\documentstyle[amssymb]{article}
\newcommand{\be}{\begin{equation}}
\newcommand{\ee}{\end{equation}}
\newcommand{\bea}{\begin{eqnarray}}
\newcommand{\eea}{\end{eqnarray}}
\newcommand{\ba}{\begin{array}}
\newcommand{\ea}{\end{array}}

\def\bbox{{\,\lower0.9pt\vbox{\hrule \hbox{\vrule height 0.2 cm\hskip
0.2 cm \vrule height 0.2 cm}\hrule}\,}} 
\newcommand{\dsl}{\pa \kern-0.5em /}

\newcommand{\nn}{\nonumber \\}

\def\ben{\begin{equation}}
\def\een{\end{equation}}
\def\bena{\begin{eqnarray}}
\def\eena{\end{eqnarray}}

\def\e{\epsilon}
\def\6{\partial}

\def\G{\Gamma}
\def\tG{\tilde \Gamma}

\def\g{\gamma}

\def\l{\lambda}

\def\mD{{\mathcal D}}

\def\today{\ifcase\month\or  January\or February\or March\or April\or
May\or June\or  July\or August\or September\or October\or November\or
December\fi \space\number\day, \number\year} 
\input amssym.def 
\input amssym.tex

\def\bE {\Bbb{E}}

\def\bft{\mbox{\boldmath $\tau$}}
\def\bfO{\mbox{\boldmath $\Omega$}}

\def\bfX{\mbox{\boldmath $X$}}
\def\bfmX{\mbox{\boldmath ${\mathcal X}$}}

\def\bfY{\mbox{\boldmath $Y$}}
\def\bfD{\mbox{\boldmath ${\mathcal D}$}}
\def\bfDel{\mbox{\boldmath $\Delta$}}

\def\bfn{\mbox{\boldmath $\nabla$}}

\def\l{\lambda}
\def\hk{$\mbox{HK}_8$}
\def\hk4{$\mbox{HK}_4$}

\begin{document}

\begin{titlepage}
\begin{flushright}
DAMTP-2003-123\\
hep-th/0311099
\end{flushright}

\vspace{4cm}

\begin{center}
{\LARGE {\bf Membrane solitons in eight-dimensional hyper-K\"ahler 
backgrounds}}
\vspace{1cm}

Rub\'en Portugues\footnote{e-mail: R.Portugues@damtp.cam.ac.uk}\\
\vspace{0.3cm}

{\it DAMTP, Centre for Mathematical Sciences, Cambridge University,\\
Wilberforce Road, Cambridge CB3 0WA, UK.}
\vspace{0.5cm}

\end{center}
\vspace{1cm}

\begin{abstract}
\noindent
We derive the BPS equations satisfied by lump solitons in $(2+1)$-dimensional
sigma models with toric 8-dimensional hyper-K\"ahler ($\mbox{HK}_8$) 
target spaces and check they preserve $\frac{1}{2}$
of the supersymmetry.  We show
how these solitons are realised in M theory as M2-branes wrapping
holomorphic 2-cycles in the $\bE^{1,2}\times \mbox{HK}_8$ background.
Using the $\kappa$-symmetry of a probe M2-brane in this background we 
determine the supersymmetry they preserve, and note
that there is a discrepancy in the fraction of supersymmetry
preserved by these solitons as viewed from the low energy effective
sigma model description of the M2-brane dynamics or the full
M theory.  Toric $\mbox{HK}_8$ manifolds are dual to a Hanany-Witten
setup of D3-branes suspended between 5-branes.  In this picture the
lumps correspond to vortices of the three dimensional ${\mathcal N}=3$ or
${\mathcal N}=4$ theory.

\end{abstract}

\end{titlepage}


\setcounter{equation}{0}

\section{Introduction}

In the absence of background field strengths, the dynamics of branes
with vanishing worldvolume gauge fields embedded in a
$(D+1)$-dimensional spacetime with coordinates $X^M$ and metric
$G_{MN}$ is governed by the Nambu-Goto action  
\be
\label{action}
S = - \int d^{p+1} \sigma \, \sqrt{-\det g}\,.
\ee
Here $\sigma^\mu$ $(\mu =0 ,\dots,p)$ denote the worldvolume
coordinates, $g$ is the pull-back of the spacetime metric onto the
worldvolume and the brane tension has been set to 1.  It is often
convenient to fix the worldvolume diffeomorphism invariance by setting
\be
X^\mu (\sigma) = \sigma^\mu\,.
\ee
We take the metric $G_{MN}$ to be the direct product of the
flat metric $\eta$ on the worldvolume and a metric $G_{AB}$
($A,B=p+1,\dots ,D$) on the Riemannian space transverse to the brane.
The induced metric becomes 
\be
g_{\mu\nu} = \eta_{\mu\nu} + \6_\mu X^A \6_\nu X^B G_{AB}\,.
\ee
Ignoring a constant term related to the mass of the brane, the
Lagrangian ${\mathcal L}$ in (\ref{action}) can be expanded to give
the sigma model Lagrangian density:
\be
{\mathcal L} \approx \frac{1}{2} \eta^{\mu\nu} \6_\mu X^A \6_\nu X^B
G_{AB} + {\mathcal O}\left( (\6 X)^4 \right)\,.
\ee
We therefore obtain the result that the low energy `non-relativistic'
fluctuations of a p-brane are governed by a $(p+1)$-dimensional sigma
model which describes the embedding of the brane in its transverse
space, where the sigma model fields $X^A$ are these transverse
coordinates\footnote{For a more detailed treatment see
\cite{Townsend:1999hi}.}.   

In \cite{Bergshoeff:1999jc}, the lumps, kinks and Q-kinks of
sigma-models with a four dimensional hyper-K\"ahler (\hk4) target
space were found to be realised in M theory as intersecting M2-branes
in a KK-monopole background and dimensional reductions of this
configuration.  In that case the sigma model describes the fluctuations of 
one of the two M2-branes and the other M2-brane appears in the 
worldvolume theory of the former as a lump-membrane.  This background
was chosen because a consistent supersymmetric sigma model can have at
most 8 supersymmetries.  The KK-monopole halves the 32
supersymmetries of M theory and then the M2-brane on which the sigma
model is defined, and which describes its vacuum, halves this again to
8.  The approach was extended in \cite{Gauntlett:2000de} and
\cite{Portugues:2002ih}, where new static and stationary sigma model
solitons were found and their corresponding realisations in M theory
studied.  

The aim of this paper is to study another class of sigma models with
relevance in M theory, namely ${\mathcal N}=4$ $(2+1)$-dimensional
sigma models with an eight dimensional hyper-K\"ahler ($\mbox{HK}_8$)
target space.  We start in section 2 by presenting a brief
introduction to toric $\mbox{HK}_8$ manifolds.  We use an explicit
form of the metric for these manifolds which includes a parameter
$\l$, such that for $\l=0$ the $\mbox{HK}_8$ manifold degenerates into
the direct product $\mbox{HK}_4\times\mbox{HK}_4$. In section 3 we
study a $(2+1)$-dimensional sigma model with an $\mbox{HK}_8$ manifold
as a target space.  We describe a consistent truncation that can be
performed on such a model that yields a 4-dimensional K\"ahler
manifold as the ``new'' target space. We obtain the first order
Bogomol'nyi equations satisfied by lumps by using the standard
procedure of writing the energy density as a sum of squares
plus a topological term.  We also check the fraction of supersymmetry
preserved by the lumps.  In section 4 we proceed to study the M theory
picture.  In order to do this we chose to examine a probe M2-brane in
a $\bE^{1,2}\times \mbox{HK}_8$ background which is a  solution to
11-dimensional supergravity.  First of all we find that for the
generic case $\l\neq 0$, the vacuum of the theory preserves only 6
supersymemtries, and therefore corresponds to an ${\mathcal N}=3$
theory in 2+1 dimensions.  This reduction in the number of
supersymmetries when compared to the sigma model was argued in
\cite{Townsend:2002wd} to be due to the higher order fermionic
interactions not captured by the sigma model\footnote{We note that an
${\mathcal N}=3$ sigma model is automatically ${\mathcal N}=4$
\cite{Alvarez-Gaume:1983ab}.}.  We then use the BPS equations derived
from the sigma model to determine the fraction of supersymmetry
preserved by the membrane lumps in M theory.  The results of the paper
are summarised in Table 1.  We note the discrepancy in the
supersymmetry preserved by the lumps as seen from the sigma model
analysis and from the probe M2-brane perspective.
\begin{table}
\begin{center}
\begin{tabular}{|l|l|c|c|} \hline
&& $\lambda=0$ & $\lambda\neq 0$ \\ \hline 
Sigma model & Vacuum & 8 & 8 \\ \cline{2-4}
 & Lump - 2 active fields & 4 & 4  \\  \cline{2-4}
& Lump - 4 active fields & 4 & 4  \\ \hline 
M theory & Vacuum & 8 & 6 \\ \cline{2-4}
& Lump - 2 active fields & 4  & no solution \\ \cline{2-4}
& Lump - 4 active fields & 2 & 2  \\ \hline 
\end{tabular}
\end{center}
\caption{We exhibit the supersymmetries of the vacuum theories and
those preserved by solutions to the lump equations in both the sigma
model and the full M theory, for the generic $\mbox{HK}_8$ manifold
$(\l\neq 0)$, and for the degenerate case when the $\mbox{HK}_8$
manifold is the direct product $\mbox{HK}_4\times\mbox{HK}_4$
($\l=0$).  } 
\end{table}

It is natural to ask why we expect the BPS equations derived from the
sigma model to be valid in M theory.  Lumps on the membrane correspond
to intersections with other membranes wrapping K\"ahler calibrated
2-cycles.  The full M theory equations describing the lump will be
those that determine such a calibrated surface.  A K\"ahler
calibration however, is fully determined by linear first order
differential equations, and we expect to be able to trust the sigma
model to this order.  For other calibrations, the sigma model agrees
only to linearized order \cite{Portugues:2002ih}. In section 5 we
conclude with a summary of these results and some comments on a
Hanany-Witten dual
version of the configurations considered here whose low energy
effective theory is a gauge theory in $(2+1)$ dimensions with either
${\mathcal N}=4$ or ${\mathcal N}=3$ depending on whether the
$\mbox{HK}_8$ manifold a direct product $\mbox{HK}_4\times\mbox{HK}_4$
or not.

\section{$\mbox{HK}_8$ manifolds}

In this paper we will consider a sigma model with a toric eight dimensional
hyper-K\"ahler target space.  These manifolds have the metric 
\be
\label{metric1}
ds^2 = U_{\alpha\beta} d\bfmX^\alpha \cdot d\bfmX^\beta +
U^{\alpha\beta} (d \varphi_\alpha + {\mathcal A}_\alpha )(d\varphi_\beta +
{\mathcal A}_\beta) \,,
\ee
where $\alpha,\beta=1,2$, $\bfmX^\alpha = (\bfX, \bfY)$, $\varphi_\alpha =
(\varphi, \psi)$, $U_{\alpha\beta}$ is a matrix of funxtions of $\bfX$
and $\bfY$ and ${\mathcal A}_\alpha=(A,B)$ is a doublet of one forms.  This
metric admits a triholomorphic $T^2$ action generated by the two
Killing vector fields $\frac{\6}{\6 \varphi}$ and $\frac{\6}{\6
\psi}$, and admits the triplet of K\"ahler two-forms
\be
\bfO = (d \varphi_\alpha + A_\alpha)d\bfmX^\alpha - \frac{1}{2}
U_{\alpha\beta} d\bfmX^\alpha \times d\bfmX^\beta \,,
\ee
where the wedge product of forms is understood.  Without loss of
generality, we take 
\be
\label{u}
U_{\alpha\beta} = \left( \begin{array} {cc} U &  W \\ W & V \end{array} \right)
\,,
\ee
where $U,V$ and $W$ are functions of the six coordinates
$(\bfX,\bfY)$.  The inverse matrix $(U^{-1})_{\alpha\beta}$ is given by
\be
\label{uinverse}
(U^{-1})_{\alpha\beta} = (UV- W^2)^{-1} \left( \begin{array} {cc} V & -
 W \\ -  W  & U \end{array} \right) \,.
\ee
The 1-forms $A$ and $B$ satisfy certain conditions analogous to the
 usual $\nabla \times A = \pm \nabla U$ of $\mbox{HK}_4$
 manifolds which ensure that (\ref{metric1}) is Ricci-flat.  These conditions
 will play no part in what follows, so we refer the reader to
 \cite{Gauntlett:1997pk} for details.
Therefore the most general form of the metric for a toric
$\mbox{HK}_8$ manifold is
\bea
\label{metric2}
ds^2 ({\mbox{HK}_8}) &= &U d\bfX \cdot d\bfX + V d\bfY \cdot d\bfY +
2  W d\bfX \cdot d\bfY \nn 
& + & (UV-  W^2)^{-1} \left( V(d\varphi + A)^2 + U(d\psi +
B )^2 - 2  W(d\varphi + A)(d\psi + B) \right)  \,.
\eea
For future use we also explicitly write down
\be
\label{2form}
\Omega^{(3)} = \mD \varphi \wedge dX_3 + \Delta \psi \wedge dY_3 - U dX_1
\wedge dX_2 - V dY_1 \wedge dY_2 -  W ( dX_1 \wedge dY_2 + dY_1
\wedge dX_2) \,,
\ee
where we have introduced the ``covariant'' derivatives
\be
\label{covder}
\mD \varphi = d \varphi + A\,, \qquad \Delta \psi = d \psi + B.
\ee
Generically, when $W\neq 0$, the manifold with metric (\ref{metric2}) will
have holonomy $Sp_{2}$.  The 32 component $SO(10,1)$ Majorana spinor
of eleven dimensions will give 6 $Sp_2$ singlets when decomposed in
terms of its $Sp_2$ subgroup, as we shall check explicitly in section
4.1.  However when $ W=0$ the $\mbox{HK}_8$ manifold becomes the
direct product $\mbox{HK}_4\times\mbox{HK}_4$, and the holonomy
is accordingly enhanced to $Sp_1 \times Sp_1$.  When this is the case,
the eleven dimensional spinor decomposes to give 8 singlets of this
subgroup.  It will be convenient in what follows to set the function
$W$ to be a non-vanishing constant, so from now on we set it to $\l$.
This ensures that the preceeding discussion applies, and allows us
to choose a gauge in which the different $U(1)$ connections $A_i$ can
be set to zero, such that the covariant derivatives become the usual
partial derivatives. 
  
\vspace{0.5cm}

\section{The sigma model approach}

\subsection{The model and the BPS equations}

We wish to study the effective ($2+1$)-dimensional sigma model
describing the fluctuations of an M2-brane.  The Lagrangian is 
\be
\label{lagrangian}
{\mathcal L} = \eta^{ij} \6_i \Phi^A \6_j \Phi^B G_{AB}\,,
\ee
where $\Phi^A = (\bfX, \varphi, \bfY, \psi)$ $(A,B=1,\dots,8)$ and
$i,j=0,1,2$ and $G_{AB}$ is given by (\ref{metric2}).  The metric
(\ref{metric2}) allows two $SO(3)$ actions, each acting
on one of the sets $(\bfX,\varphi)$ and $(\bfY,\psi)$.  In both cases
the angular variables transform in the trivial representation, whereas
$\bfX$ and $\bfY$ transform in the $\mathbf{3}$.  In the Lagrangian
(\ref{lagrangian}) however, when $\lambda\neq 0$ the term from the
metric involving $2\lambda d\bfX \cdot d\bfY$ breaks this to the
diagonal $SO(3)$.  The energy density associated to the Lagrangian
(\ref{lagrangian}) is
\bea
{\mathcal E} &= & U |{\dot{\bfX}}|^2 + U |\bfn \bfX|^2 + V |{\dot{\bfY}}|^2 + V
|\bfn \bfY|^2 +  2 \l  |{\dot{\bfX}} \cdot {\dot{\bfY}}| + 2\l  |\bfn
\bfX \cdot \bfn \bfY| \nn 
&& + (UV-\l^2 )^{-1} \bigl[ V (\mD_t \varphi)^2
+ V |\bfD \varphi |^2 + U (\Delta_t \psi)^2 + U |\bfDel \psi |^2 \nn
&& - 2 \l  (\mD_t \varphi)(\Delta_t \psi) - 2 \l ( \bfD \varphi \cdot
\bfDel \psi) \bigl]  \,,
\eea
where $\bfn=(\6_1, \6_2)$ and the covariant derivaties defined in
(\ref{covder}) have been used.  We will now choose a two-centered form
for the harmonic functions $U(\bfX)$ and $V(\bfY)$
\be
U(\bfX) = {\textstyle const.} + \frac{1}{2} \left( \frac{1}{| \bfX +
a{\hat X_3}| } + \frac{1}{| \bfX -  a{\hat X_3} | }  \right)\,,
\ee
and a similar expression for $V(\bfY)$ with two centres at $\bfY=\pm b
{\hat Y}_3$, where the hats denote unit vectors.  For this choice of
functions, and any other functions with colinear centers, the metric
(\ref{metric2}) (with $W=\l$) has two additional Killing vector fields which
generate rotations around the ${\hat X_3}$ and ${\hat Y_3}$ axes.
This breaks the $SO(3)$ isometry in the $\bfX$- and $\bfY$-planes to
$SO(2)$.  The fields will decompose into representations of these
subgroups, and in particular $\mathbf{3} \rightarrow \mathbf{2}
\oplus \mathbf{1}$.  We will now study the truncated model in which we
only keep the $SO(2)$ singlets.  Such truncations are always
consistent, that is, solutions to the truncated models will
automatically be solutions to the full model with the other fields set
to zero.  This is easy to see by considering that every term in a
Lagrangian must transform trivially under a symmetry tranformation.
Therefore any term which couples a singlet and non-singlet fields must
contain at least two of the latter. 

The truncated model has non-trivial fields $\varphi,\psi,X_3,Y_3$ and
from now on we will 
write the last two as $X$ and $Y$ for simplicity.  This model has a
$U(1)_L\times U(1)_R$ symmetry, which is again further broken to the diagonal
$U(1)$ when $\lambda\neq 0$. It constitutes a truncation to a K\"ahler
model with metric 
\be
\label{metric3}
ds^2 = U dX^2 + V dY^2 +   2\l  dX dY 
+ (UV-\l^2 )^{-1} \left( V d\varphi^2 + U d\psi^2 - 2\l 
d\varphi  d\psi  \right)  \,,
\ee
and K\"ahler two form
\be
\label{2form2}
\Omega = d \varphi \wedge dX + d \psi \wedge dY  \,.
\ee
The energy density becomes
\bea
\label{edensity}
{\mathcal E} &= &  U \left((\6_1 X)^2 +(\6_2 X)^2 \right) + V
\left((\6_1 Y)^2 +(\6_2 Y)^2 \right) + 2\l \left(\6_1 X \6_1 Y +
\6_2 X \6_2 Y \right) \nn 
&& + (UV-\l^2)^{-1} \bigl[  V \left( (\6_1\varphi)^2 +
(\6_2\varphi)^2  \right) + U \left( (\6_1 \psi)^2 + (\6_2
\psi)^2 \right)  \nn
&& - 2 \l \left( \6_1 \varphi \6_1 \psi + \6_2 \varphi
\6_2 \psi  \right)  \bigl] \,.
\eea 
Remarkably, it is now possible to write (\ref{edensity}) as
\bea
\label{wow}
{\mathcal E} & = & U \left( \6_1 X + U^{-1} \6_2 \varphi + \frac{\l}{U} \6_1
Y \right)^2 + U \left( \6_2 X - U^{-1} \6_1 \varphi + \frac{\l}{U}
\6_2 Y \right)^2 \nn 
&& + V \left( \sqrt{\frac{UV-\l^2 }{UV}} \6_1 Y + V^{-1}
\sqrt{\frac{UV}{UV-\l^2 }} \6_2 \psi - \frac{\l}{\sqrt{UV(UV-\l^2
)}} \6_2 \varphi  \right)^2 \nn
&& + V \left( \sqrt{\frac{UV-\l^2 }{UV}} \6_2 Y - V^{-1}
\sqrt{\frac{UV}{UV-\l^2 }} \6_1 \psi +
\frac{\l}{\sqrt{UV(UV-\l^2)}} \6_1 \varphi  \right)^2   \nn
&& + 2 (\6_2 X \6_1 \varphi - \6_1 X \6_2 \varphi + \6_2 Y \6_1
\psi - \6_1 Y \6_2 \psi )\,,
\eea
that is, as a sum of squares with positive coefficients plus a
topological term given by the pull-back of the K\"ahler two form
(\ref{2form2}) onto the worldvolume.  A very similar
form corresponding to interchanging $X$ and $Y$ can also be written
down, though as will become evident underneath, it will lead to the
same soliton equations.
We therefore expect the existence of solitons, whose explicit
solutions are given by the vanishing of the positive semi-definite
terms and which are supported by a topological charge, which in this
case is the pull-back of the K\"ahler 2-form (\ref{2form}) of the
truncated model.  The linear BPS equations may be written as
\bea
\label{lumpeqns}
\6_1 X & = & -\frac{V}{UV-\l^2}\, \6_2 \varphi + \frac{\l}{UV-\l^2
}\, \6_2 \psi \nn
\6_2 X & = & \frac{V}{UV-\l^2 }\,\6_1 \varphi - \frac{\l}{UV-\l^2
} \,\6_1 \psi \nn
\6_1 Y & = & -\frac{U}{UV-\l^2 }\,\6_2 \psi + \frac{\l}{UV-\l^2
} \,\6_2 \varphi \nn
\6_2 Y & = & \frac{U}{UV-\l^2 }\,\6_1 \psi - \frac{\l}{UV-\l^2
} \,\6_1 \varphi \,.
\eea
Any solution to these equations will correspond to a two-dimensional,
surface embedded in four dimensions and will be a solution to the full
eight dimensional hyper-K\"ahler sigma model. 

We also note that when $\l=0$ we obtain two sets of the usual lump
equations as written in \cite{Gauntlett:2000de}:
\bea
\label{plain}
\6_1 X = - U^{-1} \6_2\varphi & \qquad & \6_2 X = U^{-1} \6_1\varphi
\nn
\6_1 Y = - V^{-1} \6_2\psi & \qquad & \6_2 Y = V^{-1} \6_1\psi\,.
\eea

It is reassuring to check that the 2-surface described parametrically
by (\ref{lumpeqns}) is calibrated, that is to say, that its volume
form ${\mathcal V}$ is equal to the K\"ahler 2-form $\Omega$, given by
(\ref{2form2}).  To evaluate ${\mathcal V}$ on the 2-surface we simply
need to pull-back the metric (\ref{metric3}), which we denote by
$G_{IJ}$, onto the 2-surface using the lump equations
(\ref{lumpeqns}), to obtain the induced metric $\gamma_{ab}$:
\be
\gamma_{ab} = \frac{\6 X^I}{\6 \sigma^a} \frac{\6 X^J}{\6 \sigma^b}
G_{IJ}\,, 
\ee
where $X^I=(X,Y,\varphi,\psi)$.  Performing this calculation we find
that: 
\bea
&& \g_{11}=\g_{22} = |\Omega|\nn
&& \g_{12}=\g_{12} = 0 \,,
\eea
showing that 
\be
-\det(\g) = |\Omega|^2 \,.
\ee

\vspace{0.5cm}

\subsection{Sigma model supersymmetry}

As stated in the introduction supersymmetric sigma models have at most
eight supercharges.  Therefore all supersymmetric sigma models can be
obtained from a `maximal' ${\mathcal N}=1$ model in $5+1$ dimensions
via dimensional reduction.  We regard the ${\mathcal N}=4$
$(2+1)$-dimensional model (\ref{lagrangian}) we are considering as being
obtained from the maximal one by performing three trivial dimensional
reductions. 

The conditions for a bosonic solution to a sigma model to be
supersymmetric can be derived by requiring that the supersymmetry
variations of the fermions vanish.  For the target spaces
considered here, the condition was derived in \cite{Gauntlett:2000bd}
to be
\be
\label{fsusy}
[\g^m\bft \cdot \6_m \bfX^\alpha + i U^{\alpha\beta} \g^m {\mathcal D}_m
\varphi_\beta]\epsilon =0 \,, 
\ee
where the $\g^m$ are $D=6$ Dirac matrices and the spinor $\epsilon$
satisfies the chirality constraint 
\be
\label{chirality}
\g^{012345} \epsilon = \epsilon\,,
\ee
and therefore has, a priori, 8 independent real components.  
The number of solutions to (\ref{fsusy}) will be the
number of unbroken supersymmetries.  Substituting the explicit form
$U^{\alpha\beta}$ which we are considering into (\ref{fsusy}) we
obtain the pair of equations 
\bea
\label{susymain}
\biggl[ \tau^3 (\g^1 \6_1 X + \g^2 \6_2 X) + \frac{i V}{UV-\l^2}
(\g^1 \6_1 \varphi + \g^2 \6_2  \varphi)&&   \nn 
 - \frac{i \l}{UV-\l^2}
(\g^1 \6_1 \psi + \g^2 \6_2  \psi)  \biggl] \, \epsilon & = &
0 \nn
\biggl[ \tau^3 (\g^1 \6_1 Y + \g^2 \6_2 Y) + \frac{i U}{UV-\l^2 }
(\g^1 \6_1 \psi + \g^2 \6_2  \psi)&&   \nn 
 - \frac{i \l }{UV-\l^2 }
(\g^1 \6_1 \varphi + \g^2 \6_2  \varphi)  \biggl] \, \epsilon & = &
0 \,.
\eea 
Now, substituting the lump equations (\ref{lumpeqns}) into
(\ref{susymain}) implies that (\ref{susymain}) is satisfied provided
the projector
\be
\label{projector}
i \tau^3 \g^{12} \epsilon = \epsilon 
\ee
is imposed on the spinor.  This is the usual lump projector: it is
traceless and squares to the identity and as it commutes with
$\g^{012345}$, it halves the number of components of $\epsilon$.
There are therefore 4 solutions.  We note here that solutions to
(\ref{plain}) in the $\l=0$ case also preserve the same amount of
supersymmetry.

\section{M theory Supersymmetry}

In this paper we consider an M2-brane as a probe brane in the
spacetime 
\be
\label{11dmetric}
ds^2_{11d} = ds^2 (\bE^{(1,2)}) + ds^2 ({\mbox{HK}_8})\,,
\ee
which is a Ricci flat solution of eleven dimensional supergravity,
where $ds^2 ({\mbox{HK}_8})$ is the metric (\ref{metric2}) with
$W=\lambda$. 

Let us now consider a (bosonic) probe brane in this background. The
global fermionic symmetries of the background will act on the membrane
fields, and in general a solution will not be left invariant by these
transformations.
The condition for the brane to be supersymmetric is that there should
exist a $\kappa$-symmetry transformation that combines with the global
fermionic symmetries to leave the fields invariant
\cite{Bergshoeff:1987dh}, \cite{Becker:1995kb},
\cite{Bergshoeff:1997kr}. 

In this approach the brane does not backreact.  It was
noted in \cite{Bergshoeff:1997kr} that considering backreacting branes
does not impose any additional constraints on supersymmetry, that is,
that a spinor will satisfy the Killing spinor equation if and only if
it generates an appropiate $\kappa$-symmetry transformation on the
worldvolume, as described above.

\subsection{Covariantly constant spinors in toric $\mbox{HK}_8$ manifolds}

Before proceeding we need to determine the conditions imposed on the
covariantly constant spinors of the metric (\ref{11dmetric}).  As
explained in the appendix, where we also describe the gamma matrix
conventions we use, we write an $SO(10,1)$ Majorana spinor $\e$ as
\be
\epsilon = \chi(\sigma^i) \otimes \eta(X^A) \,,
\ee
where $\chi$ is an anticommuting, $SO(2,1)$ spinor,
and $\eta$ is a commuting,  $SO(8)$ spinor.  The
condition that $\epsilon$ should be covariantly constant with respect
to the metric (\ref{11dmetric}) implies that $\eta$ should be
covariantly constant on the $\mbox{HK}_8$ manifold
\cite{Gauntlett:1997pk} \cite{Becker:1996gj}.  We introduce a vielbein
\bea
e^{X_i} & = & U^{\frac{1}{2}}(dX^i + \l U^{-1} \delta_{i \alpha}
dY^\alpha) \nn
e^\varphi & = &  \left(\frac{V}{K}\right)^{\frac{1}{2}} \left[ (d\varphi + A_i
dX^i ) - \l V^{-1} (d\psi + B_\alpha dY^\alpha) \right] \nn
e^{Y_\alpha} & = & \left( \frac{K}{U} \right)^{\frac{1}{2}} dY^\alpha \nn
e^\psi & = & V^{-\frac{1}{2}} (d\psi + B_\alpha dY^\alpha) \,,
\eea
where $i=1,2,3$ labels the $X$ coordinates, $\alpha=1,2,3$ labels the
$Y$ coordinates and $K=UV-\l^2$.  Killing spinors of (\ref{metric2})
will be independent of the the coordinates $\varphi$ and $\psi$
\cite{Gauntlett:1997pk}.  This gives us the following conditions on
the Killing spinors:
\bea
\label{si}
\frac{1}{\sqrt{UV}}\left[ K \,\tG_{\varphi X_1 X_2 X_3} + \lambda
\sqrt{UV} \,\tG_{X_1 Y_1 } -\lambda \,\tG_{X_1 \varphi } ( {\sqrt K}
\,\tG_{X_3 Y_2} + \l\, \tG_{Y_2 Y_3 } - {\sqrt K} \,\tG_{X_2 Y_3}
)\right] \eta & = & \eta \nn 
\left[ \sqrt{\frac{UV}{K}} \tG_{\psi Y_1 Y_2 Y_3} + \frac{\l}{\sqrt K}
\tG_{\psi\varphi} \right] \eta & = & \eta \,. \nn
\eea
The explicit form of the Killing spinor can be determined by looking
at the other components of the covariantly constant spinor equation,
and we do so here for completeness.  For $\l=0$ the Killing spinor is
in fact constant.  For $\l\neq 0$ we obtain from $\nabla_{X_i} \eta
=0$ the condition
\be
\eta = \exp\left[ -2 \tan^{-1} \left( \frac{\sqrt K}{\lambda}\right)
\left( \tG_{X_1 Y_1} + \tG_{X_2 Y_2} + \tG_{X_3 Y_3} \right)\right] \,
\eta_0\,,
\ee
whereas from the covariant derivative $\nabla_{Y_\alpha} \eta =0$ we
obtain 
\be
\eta = \exp\left[ -2 \tan^{-1} \left( \frac{\sqrt K}{\lambda}\right)
 \tG_{\varphi \psi } \right] \,
\eta_0\,.
\ee
These are consistent if and only if
\be
\left( \tG_{X_1 Y_1} + \tG_{X_2 Y_2} + \tG_{X_3 Y_3} - \tG_{\varphi \psi }
\right) \eta_0 = 0 \,.
\ee
This condition must be imposed on $\eta_0$ apart from those arising
from (\ref{si}).

We can now determine the number of independent components of the
Killing spinor preserved by (\ref{si}).
First of all we note that when
$\l=0$ (and $K=UV$) these simply yield $\tG_{\psi Y_1
Y_2 Y_3} \eta = \tG_{\varphi X_1 X_2 X_3} \eta = \eta$, which commute
and therefore preserve 4 of the sixteen supersymmetries in 8
dimensions.  To analyse (\ref{si}) for $\l\neq 0$ we can choose a
representation for the $SO(8)$ Dirac matrices as explained in the
Appendix. It is then possible to check that the conditions (\ref{si})
preserve three independent components of $\eta$ in the generic $\l\neq
0$ case, as expected for a manifold of $Sp_2$ holonomy.  We label
these three independent solutions $\eta_{(i)}$ where $i=1,2,3$.  A
general covariantly constant spinor of the metric (\ref{11dmetric})
can therefore be written as
\be
\label{genspinor}
\e = \sum_i \sum _s a_{si}(\chi_s \otimes \eta_{(i)})\,,
\ee
where $s=+,-$, $\chi_+ = \left(\begin{array}{c} 1 \\ 0 \end{array}
\right)$, $\chi_- = \left(\begin{array}{c} 0 \\ 1 \end{array}
\right)$  and $a_{si}$ are six arbitrary constants.

\subsection{Supersymmetry analysis}

As described at the beginning of the section and explained in
\cite{Bergshoeff:1987dh}, \cite{Becker:1995kb},
\cite{Bergshoeff:1997kr}, the M theory condition for a bosonic
M2-brane solution to be supersymmetric is that there exists a
worldvolume $\kappa$-symmetry transformation of the form
\be
\label{msusy}
\G_\kappa \epsilon = \epsilon \,,
\ee
where $\e$ is a covariantly constant spinor of the background
normalised such that $\e^T\e=1$, and where 
\be
\Gamma_\kappa = \frac{1}{6\sqrt {-g}} \epsilon^{ijk} \6_i X^M \6_j X^N
\6_k X^P  \G_{MNP}\,.
\ee
The dimension of the space of solutions to (\ref{msusy}) is then equal
to the number of unbroken supersymmetries.
The conventions for the gamma matrices are explained in the appendix.
After choosing the physical gauge we can write (\ref{msusy}) as 
\be
\label{full}
\sqrt {-g} \epsilon = \left( 1 - \G^i\6_i X^A \G_A - 
\frac{1}{2} \G^{ij} \6_i X^A \6_j X^B  \G_{AB} \right)
\G_\star\epsilon  \,,
\ee
where $\G_\star = \G_{012}$.  Equating powers in $\6 X$, the zeroth
order condition 
\be
\label{zeroth}
\G_\star \, \e = \e 
\ee
can be rewritten using (\ref{tensor1}) and (\ref{tensor2}) as
\be
\left( \g_{012} \otimes \tG_9 \right)(\chi\otimes \eta) = (\chi\otimes
\eta) 
\ee
where we see that $\g_{012}\chi=\chi$, which is trivial, and that $\eta$
is chiral in the eight-dimensional sense and has therefore eight
independent real components.  This condition defines the vacuum of the
theory.  As it happens, the chirality constraint on $\eta$ is
automatically satisfied by solutions to (\ref{si}) as can be checked
again by using the explicit representation mentioned above.  Therefore,
(\ref{si}) are the conditions which determine the supersymmetry
preserved by the vacuum of the theory, namely it preserves
$\frac{3}{16}$ of the 32 M theory supersymmetries.

To first order equation (\ref{full}) reads 
\be
\label{first}
\G^i\6_i X^A \G_A \, \epsilon = 0 \,.
\ee
Before we study the implications of (\ref{first}), we will show that
this condition implies that the full equation (\ref{full}) is
satisfied.  By iterating (\ref{first}) we obtain the relations
\bea
\G^{ij}\6_i X^A \6_j X^B \G_{AB} \, \e &=&  - \eta^{ij} \tilde g_{ij}
\, \e
\nn 
\G^{ijk}\6_i X^A \6_j X^B \6_k X^C \G_{ABC} \, \e &=&  0 \nn 
2 \eta^{ij}\eta^{kp} \tilde g_{ik} \tilde g_{jp} \, \e&=&  (\eta^{ij}
\tilde g_{ij})^2 \, \e \,,
\eea
where we have defined $\tilde g_{ij} = \6_i X^I \6_j X^J g_{IJ} $.
These relations and the expansion
\be
\det({-\tilde g}) = 1 + \eta^{ij} \tilde g_{ij} + \frac{1}{2} (\eta^{ij}
\tilde g_{ij} )^2 - \frac{1}{2} \eta^{ij}\eta^{kp} \tilde g_{ik}
\tilde g_{jp} \,
\ee
for the determinant appearing in (\ref{full}) are all that are needed
to establish that studying the condition (\ref{first}) is sufficient to
ensure supersymmetry.  This is the statement that the equations
defining a K\"ahler calibration are linear and first order in
derivatives.  This is not true for other calibrations \cite{hl},
although equation (\ref{first}) will still reproduce these to first
order \cite{Portugues:2002ih}.

\vspace{.5truecm}

We can now proceed to understand the implications of (\ref{first}).
To do so, let us introduce a vierbein for the metric (\ref{metric3})
\bea
\label{vierbein}
e^X&=& U^{1/2} \left( dX + \frac{\l }{U} dY \right) \nn  
e^Y&=&  \left( \frac{UV-\l^2 }{U}\right)^{1/2} dY  \nn
e^\varphi &=& \left( \frac{V}{UV-\l^2 }\right)^{1/2} \left(
d\varphi - \frac{\l}{V} d\psi \right)  \nn
e^\psi &=& V^{-1/2}  d\psi \,, 
\eea
and use it to write the
curved Dirac matrices in terms of the tangent space ones $\G_A=(\G_{X_3},
\G_{Y_3}, \G_\varphi, \G_\psi)$, where we have introduced the
subscript 3 to make a direct comparison with section 4.1.  Using this
notation and the new lump equations (\ref{lumpeqns}), the condition
(\ref{first}) implies the following two projectors
\bea
\left(  \l \G^2 \G_\varphi + \G^1 \sqrt{UV} \G_{Y_3} - \G^2 \sqrt{UV-\l^2
}  \G_\psi \right )\,\epsilon &=& 0 \nn
\left(  \l  \G^2 \G_\psi - \G^1 \sqrt{UV} \G_{X_3} + \G^2 \sqrt{UV-\l^2
}  \G_\varphi \right )\,\epsilon &=& 0 \,.
\eea
We can write these succintly as
\bea
\label{projectors}
\left(\G^{12}\G_{Y_3 \psi} \cosh \tau + \G_{\psi\varphi} \sinh\tau
\right)\,\e & = & \e \nn
\left(\G^{12}\G_{X_3 \varphi} \cosh \tau + \G_{\psi\varphi}\sinh\tau
\right)\,\e & = & \e \,,
\eea
where the parameter $\tau$ is defined by 
\be
\label{tau}
\cosh\tau = \frac{\sqrt{UV}}{\sqrt{UV-\l^2}} \qquad \sinh\tau =
\frac{\l}{\sqrt{UV-\l^2}} \,.
\ee
It is now possible to use the explicit representation (\ref{tensor2})
to impose the conditions (\ref{projectors}) on the general form of the
covariantly constant spinor (\ref{genspinor}).  On doing this it is
easy to check that only two of the arbitrary constants $a_{si}$ remain
independent, signifying that general solutions preserve two
supersymmetries.  We may ask ourselves whether a subset of the
solutions to the equations (\ref{lumpeqns}), for example, whether
setting $Y=0$ identically, may preserve more supersymmetry, but it can
be checked that this is not the case.

It is possible to run through the same argument in the $\l=0$ case.
In this case the equations satisfied by the lumps are (\ref{plain})
and the projector (\ref{first}) imposes the conditions 
\bea
\label{projectorsplain}
\G^{12}\G_{Y_3 \psi} \,\e & = & \e \nn
\G^{12}\G_{X_3 \varphi} \,\e & = & \e \,.
\eea
These commute between themselves and with the other conditions imposed
on the covariantly constant spinors in this case.  We note that we can
have solutions where only one of the pairs $(X_3,\varphi)$ or
$(Y_3,\psi)$ is non-zero.  This would be a solutions with two active
fields.  From the point of view of the sigma model, all solutions,
with either two or four active fields, preserve four supersymmetries.
This is not the case in the context of M theory, where only solutions
with two active fields preserve four supersymmetries and solutions
with four active fields preserve only two. 

A summary of the results is presented in Table 1 displayed in the
Introduction.

\section{Conclusions and Discussions}

We summarise the results of the paper as follows.  We studied a
$(2+1)$-dimensional sigma model with an $\mbox{HK}_8$ target space and
derived the BPS equations that must be satisfied by lump solitons.  The
sigma model seemed to suggest that solutions to these equations
preserve 4 supersymmetries. We then analysed the membranes as probes in
the M theory background (\ref{11dmetric}).  We checked explicitly that the
vacuum of the theory generically has less supersymmetries than the
sigma model, and that solutions to the lump equations, which we expect
to hold in M theory preserve only two supersymmetries.

The setup considered here of an M2-brane in the background
(\ref{11dmetric}) is dual to a Hanany-Witten configuration
\cite{Hanany:1996ie} in which the D3-brane suspended between an NS5-brane
and a $(p,q)$-5-brane \cite{Kitao:1998mf,Gukov:2002er} (the choice of $W=\l$ is
constant is necessary for this duality).  The effective field theory
on the D3-brane is a $(2,1)$-dimensional gauge theory.  When the
$\mbox{HK}_8 = \mbox{HK}_4 \times \mbox{HK}_4 $ this dual theory will
also have ${\mathcal N}=4$ supersymmetry, but for generic $\mbox{HK}_8$, that
is, when $\l\neq 0$, this theory will include a Chern-Simons term that
will preserve only ${\mathcal N}=3$.  In fact, we expect the lumps
studied above to correspond to vortices in such a model, and
following the results in this paper, we expect such vortices to
preserve $\frac{1}{3}$ of the 6 supercharges, in agreement
with \cite{Lee:1999ze,Kitao:1999uj}.

\vspace{.5truecm}

\noindent
{\bf Acknowledgements}:  The author would like to thank Tibra Ali,
Sean Hartnoll, Carlos N\'{u}\~{n}ez, Guillermo Silva, David Tong and
especially Paul Townsend for useful discussions and Trinity College,
Cambridge for financial support. 

\appendix
\section{Gamma matrix conventions}

Throughout we take the 11-dimensional Dirac matrices to satisfy the
Clifford algebra
\be
\lbrace \G_M,\G_N \rbrace = 2 \, G_{MN}\,.
\ee
Following the decomposition
\be
\label{tensor1}
SO(10,1)\rightarrow SO(2,1) \times SO(8)
\ee
we can write these in tensor product form as 
\be
\label{tensor2}
\G_M = (\G_i,\G_A) = (\g_i \otimes \tG^9  , 1 \otimes \tG_A)\,,
\ee
where $M=0,\dots,9,\natural$ ($\natural$ denotes the 11-th dimension),
$i=0,1,2$ are the worldvolume directions and $A=3,\dots,9,\natural$
the $\mbox{HK}_8$ directions.  The $\g_i$ are a representation of the
$SO(2,1)$ matrices given explicitly by
\be
\g_0 = \left( \begin{array}{cc} 0 & 1 \\ -1 & 0 \end{array}
\right) \qquad 
\g_1 = \left( \begin{array}{cc} 0 & 1 \\ 1 & 0 \end{array}
\right) \qquad 
\g_2 = \left( \begin{array}{cc} 1 & 0 \\ 0 & -1 \end{array}
\right) \,,
\ee
and the $\tG$ are a representation of the $SO(8)$ gamma matrices.  We
take this to be the representation described in an appendix to Chapter 5 of
\cite{Green:sp} which is real.  $\tG^9$ is the chirality matrix in
eight dimensions, defined by $\tG^9=\tG_{[3\dots}\tG_{\natural]}$ and
satisfying $(\tG^9)^2 = 1$ and $\lbrace \tG^9 , \tG_A \rbrace =
0$. This ensures that the product of all the 11 dimensional Dirac
matrices is the identity.     

Following (\ref{tensor1}) we can write an
11-dimensional spinor $\epsilon$ as
\be
\epsilon = \chi(\sigma^i) \otimes \eta(X^A) \,.
\ee
The charge conjugation matrix in the 11-dimensional representation
(\ref{tensor2}) is given by $\G_0$.  As usual the spinor $\epsilon$
satisfies a Majorana condition in 11 dimensions which requires it to
be real and therefore have 32 independent real components.

\end{document}